\documentclass[prd,aps,nofootinbib,showpacs,12pt]{revtex4}
\usepackage{amsmath}

\begin{document}

\title{ What can we learn from $B\to a_1(1260)(b_1(1235))\pi(K)$ decays?}
\author{ Wei Wang$^{a}$, Run-Hui Li$^{b,a}$ and Cai-Dian L\"u$^{a,c}$ }
\affiliation{
 $^a$ Institute of High Energy Physics, Chinese Academy
 of Sciences, Beijing 100049, P.R. China\\
 $^b$ Physics Department, Shandong University, Jinan 250100, P.R. China\\
 $^c$ Theoretical Physics Center for Science Facilities,  CAS, Beijing 100049, P.R. China}

\begin{abstract}
We investigate the $B\to  a_1(1260)(b_1(1235))\pi(K)$ decays under
the factorization scheme and find many discrepancies between
theoretical predictions and the experimental data. In the tree
dominated processes, large contributions from color-suppressed tree
diagrams are required in order to accommodate with the large decay
rates of $B^-\to a_1^0\pi^-$ and $B^-\to a_1^-\pi^0$. For $\bar
B^0\to (a_1^+, b_1^+)K^-$ decays which are both induced by $b\to s$
transition, theoretical predictions on their decay rates are larger
than the data by a factor of 2.8 and 5.5, respectively. Large
electro-weak penguins or some new mechanism are expected to explain
the branching ratios of $B^-\to b_1^0K^-$ and $B^-\to a_1^-\bar
K^0$. The soft-collinear-effective-theory has the potential to
explain large decay rates of $B^-\to a_1^0\pi^-$ and $B^-\to
a_1^-\pi^0$ via a large hard-scattering form factor $\zeta_J^{B\to
a_1}$. We will also show that, with proper charming penguins,
predictions on the branching ratios of $\bar B^0\to (a_1^+,
b_1^+)K^-$ can also be consistent with the data.
\end{abstract}


\maketitle

\section{Introduction}

Since the first measurement on $B^0/\bar B^0\to
a_1^\mp(1260)\pi^\pm$ decays reported by BaBar and Belle
collaborations~\cite{Aubert:2006dd,:2007jn,Aubert:2006gb}, many
charmless $B$ decays into a pseudo-scalar and an axial-vector meson
have been observed. Among the 18 $B\to a_1(1260)(b_1(1235)\pi(K)$
\footnote{In the following, we will use $a_1(b_1)$ to denote the
$a_1(1260)(b_1(1235))$ meson for simplicity.} decay channels, 10 of
them have been measured with large branching ratios. Besides decay
rates,  direct CP asymmetries in some $B\to (a_1,b_1)K$ channels and
time-dependent CP asymmetries in $B^0/\bar B^0\to a_1^\pm\pi^\mp$
and $B^0/\bar B^0\to b_1^\pm\pi^\mp$ were also studied in the two B
factories~\cite{Palombo:2007jj,Aubert:2007xd,:2007kp,Aubert:2007ds,Mohanty:2007pj,:2008eq}.
Without any doubt, these results are helpful to investigate
production mechanisms of axial-vectors in $B$ decays, extract
hadronic parameters such as strong phases in $B\to AP$ decays and
probe the structures of axial-vectors.

Charmless two-body $B\to AP$ decays have received considerable
theoretical
efforts~\cite{Chen:2005cx,Calderon:2007nw,Laporta:2006uf,Yang:2007sb,Cheng:2007mx}.
Among these predictions, many of them are not consistent with each
other: most predictions by Calderon, Munoz and
Vera~\cite{Calderon:2007nw} are larger than predictions given by
Laporta, Nardulli and Pham~\cite{Laporta:2006uf} and   the QCD
factorization (QCDF) approach. Predictions on $B\to a_1\pi$ by
Laporta, Nardulli and Pham (using the second sets of form factors)
are very close to results in the QCDF approach. However there are
large discrepancies in other predictions (See
Ref.~\cite{Cheng:2007mx} for a detailed comparison between these
theoretical predictions). Many results of the QCDF approach agree
with the experimental data, but there still exist some deviations.

In the present paper, we intend to analyze the 18 $B\to AP$ decays
with the help of experimental data. We  try to check whether these
problems can be removed in the perturbative QCD (PQCD) approach and
the soft-collinear-effective-theory (SCET). Another objective is to
extract the $B\to A$ form factors through $\bar B^0\to a_1^+\pi^-$
and $\bar B^0\to b_1^\pm\pi^\mp$ decays.

\section{naive factorization approach}

The effective Hamiltonian describing $b\to D (D=d,s)$ transitions
are given by~\cite{Buchalla:1995vs}:
 \begin{eqnarray}
 {\cal H}_{eff} &=& \frac{G_{F}}{\sqrt{2}}
     \bigg\{ \sum\limits_{q=u,c} V_{qb} V_{qD}^{*} \left[
     C_{1}  O^{q}_{1}
  +  C_{2}  O^{q}_{2}+{\sum\limits_{i=3}^{10}} C_{i}  O_{i} \right]\bigg\}+ \mbox{H.c.} ,
 \label{eq:hamiltonian}
\end{eqnarray}
where $V_{qb(D)}$ are the Cabibbo-Kobayashi-Maskawa (CKM) matrix
elements. Functions $O_{i}$ are the local four-quark operators,
while functions $C_i$ are the corresponding Wilson coefficients.  It
is convenient to define combinations $a_i$ of the Wilson
coefficients:
\begin{eqnarray}\label{eq:combination}
  &&a_1= C_2+C_1/3,  \;\;a_2= C_1+C_2/3, \nonumber \\
  &&a_i=C_i+C_{i+1}/N_c\;\; (i=3,5,7,9),\nonumber \\
  &&a_i=C_i+C_{i-1}/N_c\;\; (i=4,6,8,10).
\end{eqnarray}
There exist a hierarchy for the Wilson coefficients:
\begin{eqnarray}
 a_1\gg {\rm max}[ a_2,a_{3-10}].
\end{eqnarray}

For tree-dominated processes $B^0/\bar B^0\to a_1^\pm \pi^\mp$, the
factorization formulae can be written as:
\begin{eqnarray}
 {\cal A}(\bar B^0\to a_1^+\pi^-)&=& \frac{G_F}{\sqrt 2}m_B^2 f_\pi
 V_0^{B\to a_1}\left\{V_{ub}V_{ud}^* [a_1+a_4+a_{10}+r_\pi
 (a_6+a_8)]\right.\nonumber\\
 &&\;\;\;\; \left.+V_{cb}V_{cd}^* [a_4+a_{10}+r_\pi (a_6+a_8)]\right\},\\
 {\cal A}(\bar B^0\to \pi^+a_1^-)&=& \frac{G_F}{\sqrt 2}m_B^2
 f_{a_1}
 f_+^{B\to \pi}\left\{V_{ub}V_{ud}^* [a_1+a_4+a_{10}]
 +V_{cb}V_{cd}^* [a_4+a_{10}]\right\},
\end{eqnarray}
where $r_\pi=2m_0^\pi/m_B$ with $m_0^\pi$  the chiral scale
parameter for pion. The CKM matrix elements for tree operators
$|V_{ub}V_{ud}^*|\sim 4\times 10^{-3}$ have the same order magnitude
with those for penguin operators  $|V_{cb}V_{cd}^*|\sim 8\times
10^{-3}$. Because of the hierarchy in the Wilson coefficients,
penguin contributions from the operators $O_{3-10}$ are small
compared with those from tree operators. Thus penguin contributions
can be neglected in the study of branching ratios (but crucial to CP
asymmetries). Combined with the $\bar B^0\to\pi^+\pi^-$
data~\cite{Barberio:2007cr}
\begin{eqnarray}
 {\cal BR}(\bar B^0\to \pi^+\pi^-)=(5.16\pm 0.22)\times 10^{-6},
\end{eqnarray}
we arrive at the $a_1$ meson decay constant and $B\to a_1$ form
factor:
\begin{eqnarray}
 f_{a_1}= [2.02\pm0.26\pm0.04+{\cal O}(\frac{a_{3-10}}{a_1})] f_\pi,\;\;\;
 V_0^{B\to a_1}=(1.55\pm0.28\pm0.03+{\cal O}(\frac{a_{3-10}}{a_1})]) f_+^{B\to
 \pi},
\end{eqnarray}
where the uncertainties are from  the experimental results for
branching ratios. As a rough estimation, we take $f_\pi =131$ MeV
and $f_+^{B\to \pi}=0.25$ which corresponds to
$f_{a_1}=(265\pm34\pm6)$ MeV and $ V_0^{B\to
a_1}=0.39\pm0.07\pm0.01$. These results are well consistent with
predictions based on the PQCD approaches~\cite{Wang:2007an} and
light-cone sum rules~\cite{Yang:,Wang:2008bw}.

Now we come to the two channels $B^-\to a_1^0\pi^-$ and $B^-\to
a_1^-\pi^0$ whose factorization formulae are given by:
\begin{eqnarray}
 \sqrt2{\cal A}( B^-\to \pi^- a_1^0)&=& \frac{G_F}{\sqrt 2}m_B^2 f_\pi
 V_0^{B\to a_1}\left\{V_{ub}V_{ud}^* [a_1+a_4+a_{10}+r_\pi (a_6+a_8)]\right.\nonumber\\
 &&\;\;\;\;\;\;\; \left.
 +V_{cb}V_{cd}^* [a_4+a_{10}+r_\pi (a_6+a_8)]\right\} \nonumber\\
 &+&\frac{G_F}{\sqrt 2}m_B^2 f_{a_1}
 f_+^{B\to \pi}\left\{V_{ub}V_{ud}^* [a_2-a_4+\frac{1}{2}a_{10}]
 +V_{cb}V_{cd}^* [-a_4+\frac{1}{2}a_{10}]\right\}, \\
 \sqrt 2{\cal A}( B^-\to \pi^0 a_1^-)&=& \frac{G_F}{\sqrt 2}m_B^2 f_\pi
 V_0^{B\to a_1}\left\{V_{ub}V_{ud}^* [a_2-a_4+\frac{1}{2}a_{10}+r_\pi (-a_6+\frac{1}{2}a_8)]\right.\nonumber\\
 &&\;\;\;\;\;\;\; \left.
 +V_{cb}V_{cd}^* [-a_4+\frac{1}{2}a_{10}+r_\pi (-a_6+\frac{1}{2}a_8)]\right\} \nonumber\\
 &+&\frac{G_F}{\sqrt 2}m_B^2 f_{a_1}
 f_+^{B\to \pi}\left\{V_{ub}V_{ud}^* [a_1+a_4+a_{10}]
 +V_{cb}V_{cd}^* [a_4+a_{10}]\right\}.
\end{eqnarray}
Because of the small values of $a_{3-10}$, the penguin contributions
can be safely neglected:
\begin{eqnarray}
 \sqrt2{\cal A}( B^-\to \pi^- a_1^0)&=& \frac{G_F}{\sqrt 2}m_B^2 V_{ub}V_{ud}^* [a_1f_\pi
 V_0^{B\to a_1}+a_2f_{a_1}
 f_+^{B\to \pi}],\\
 \sqrt 2{\cal A}( B^-\to \pi^0 a_1^-)&=& \frac{G_F}{\sqrt 2}m_B^2 V_{ub}V_{ud}^* [a_2f_\pi
 V_0^{B\to a_1}+a_1f_{a_1} f_+^{B\to \pi}].
\end{eqnarray}
Furthermore, in the hierarchy of $a_2\ll a_1$, branching ratios are
required to satisfy the following relation:
\begin{eqnarray}
 {\cal BR}(\bar B^0\to a_1^+\pi^-)=2{\cal
 BR}(B^-\to\pi^- a_1^0),\;\;\;
 {\cal BR}(\bar B^0\to\pi^+ a_1^-)=2{\cal BR}(B^-\to a_1^-\pi^0).
\end{eqnarray}
But the experimental data shows:
\begin{eqnarray}
 {\cal BR}( B^-\to \pi^0a_1^-)>{\cal BR}(\bar B^0\to
 a_1^-\pi^+),\;\;\;
 {\cal BR}( B^-\to \pi^-a_1^0)>{\cal BR}(\bar B^0\to
 a_1^+\pi^-),
\end{eqnarray}
which is dramatically different.  This situation is very similar
with that in $B\to\pi\pi$ decays: the branching ratio of $ B^-\to
\pi^0\pi^-$ is measured with almost equal magnitude with ${\cal
BR}(\bar B^0\to \pi^-\pi^+)$ but it is expected as one half of
${\cal BR}(\bar B^0\to \pi^-\pi^+)$ . To solve these problems, an
efficient way is to enhance the color-suppressed contribution which
is proportional to $a_2$. For example, if the Wilson coefficient
$a_2$ can be enhanced to $0.5$, the branching ratios of ${\cal BR}(
B^-\to \pi^0a_1^-)$ and ${\cal BR}( B^-\to \pi^-a_1^0)$ are
predicted as $20.0\times 10^{-6}$ and $16.7\times 10^{-6}$, where we
have utilized the experimental data on branching ratios of $B^0/\bar
B^0\to\pi^\pm a_1^\mp$. And these results are well consistent with
the experimental data.

The decay constant of $b_1$ vanishes because of the G-parity, thus
$\bar B^0\to \pi^+ b_1^-$ is factorization-suppressed and   only
  the $\bar B^0\to\pi^- b_1^+$ decay survives. From the
experimental results collected in table~\ref{tab:BR}, we can infer
that the form factors of $B\to a_1$ and $B\to b_1$ are almost equal
in magnitude at maximally recoiling: $|V_0^{B\to a_1}(q^2=0)|\simeq
|V_0^{B\to b_1}(q^2=0)|\simeq 0.35$.  One should be careful that the
two form factors have different signs, if we use LCDAs of $a_1$ and
$b_1$ evaluated by the QCD sum rules. The absolute value of these
form factors can be checked by the future measurements on
semi-leptonic $B\to A$ decays such as $\bar B^0\to
(a_1^+,b_1^+)l^-\bar\nu$.

Flavor structures of $\bar B^0\to b_1^+K^-$ and $\bar B^0\to
a_1^+K^-$ are the same with each other, thus they have the same
factorization formulae:
\begin{eqnarray}
 {\cal A}(\bar B^0\to (a_1^+,b_1^+) K^-)&=& \frac{G_F}{\sqrt 2}m_B^2 f_K
 V_0^{B\to (a_1,b_1)}\left\{V_{ub}V_{us}^* [a_1+ a_4+a_{10}+r_K (a_6+a_8)]\right.\nonumber\\
 &&\;\;\;\; \left.
 +V_{cb}V_{cs}^* [a_4+a_{10}+r_K (a_6+a_8)]\right\}.
\end{eqnarray}
The same Wilson coefficients and almost equal form factors will
induce almost equal branching ratios for $\bar B^0\to a_1^+K^-$ and
$\bar B^0\to b_1^+K^-$. To reduce the  uncertainties, we will
utilize the $\bar B^0\to \pi^+ K^-$ decay which also has the same
flavor structures with $\bar B^0\to (a_1^+,b_1^+)K^-$. The only
difference between the three modes is the different form factors
which can be extracted from tree-dominated processes $\bar
B^0\to\pi^+\pi^-$ and $\bar B^0\to (a_1^+,b_1^+)\pi^-$ decays. The
branching ratio of $\bar B^0\to \pi^+ K^-$ has been measured
as~\cite{Barberio:2007cr}:
\begin{eqnarray}
{\cal BR}(\bar B^0\to \pi^+ K^-)=(19.4\pm0.6)\times 10^{-6},
\end{eqnarray}
which implies:
\begin{eqnarray}
 {\cal BR}(\bar B^0\to a_1^+ K^-)&=&
 45.9\times 10^{-6},\;\;\;\;
 {\cal BR}(\bar B^0\to b_1^+K^-)=
 41.0\times 10^{-6}.
\end{eqnarray}
Comparing with the experimental measurements in table~\ref{tab:BR},
we see that our theoretical prediction on ${\cal BR}(\bar B^0\to
a_1^+K^-)$ is 2.8 times larger while the prediction on ${\cal
BR}(\bar B^0\to b_1^+K^-)$ is  5.5 times larger. This discrepancy
should be clarified by the theoretical studies with next-to-leading
order corrections and improved experimental measurements.

Besides $\bar B^0\to (a_1^+,b_1^+)K^-$ decays,  $B^-\to a_1^-\bar
K^0$ and $B^-\to b_1^0K^-$ decays are also measured by
experimentalists whose factorization formulae are:
\begin{eqnarray}
 {\cal A}(B^-\to a_1^-\bar K^0)&=& \frac{G_F}{\sqrt 2}m_B^2 f_K
 V_0^{B\to a_1}\left\{V_{ub}V_{us}^* [a_4-\frac{1}{2}a_{10}+r_K (a_6-\frac{1}{2}a_8)]\right.\nonumber\\
 &&\;\;\;\; \left.
 +V_{cb}V_{cs}^* [a_4-\frac{1}{2}a_{10}+r_K (a_6-\frac{1}{2}a_8)]\right\},\\
 \sqrt2{\cal A}(B^-\to b_1^0K^-)&=&  \frac{G_F}{\sqrt 2}m_B^2 f_K
 V_0^{B\to b_1}\left\{V_{ub}V_{us}^* [a_1+ a_4+a_{10}+r_K (a_6+a_8)]\right.\nonumber\\
 &&\;\;\;\; \left.
 +V_{cb}V_{cs}^* [a_4+a_{10}+r_K (a_6+a_8)]\right\}.
\end{eqnarray}
In these $b\to s$ transitions, the CKM matrix elements for penguin
operators are $|V_{cb}V_{cs}^*|\sim 40\times 10^{-3}$ and those for
tree operators are $|V_{ub}V_{us}^*|\sim 0.8\times 10^{-3}$.
Recalling the values for the Wilson coefficient combinations:
$a_1\sim 1$ and $a_4\sim a_6\sim -0.03$, we can see that
contributions from tree operators with the coefficient $a_1$ are
smaller than that from penguin operators at least by a factor of $2$
in magnitude. In order to characterize the contribution from tree
operators and symmetry breaking effects between $B^-$ and $\bar B^0$
mesons, it is useful to define the two ratios:
\begin{eqnarray}\label{eq:ratios}
 R_1\equiv\frac{{\cal BR}(B^-\to a_1^-\bar K^0)}{{\cal BR}(\bar B^0\to a_1^+K^-)}\times
 \frac{\tau_{\bar B^0}}{\tau_B^-},\;\;\;
 R_2\equiv\frac{{\cal BR}(B^-\to b_1^0K^-)}{{\cal BR}(\bar B^0\to
 b_1^+K^-)}\times \frac{\tau_{\bar B^0}}{\tau_B^-},
\end{eqnarray}
where $\tau$ is the lifetime of $B$ meson.  Neglecting tree
operators and electro-weak penguins, the ratios obey the limit:
\begin{eqnarray}\label{eq:ratiosthe}
 R_1=1,\;\;\; R_2=0.5,
\end{eqnarray}
which are quite different from the experimental results:
\begin{eqnarray}\label{eq:ratiosexp}
 R_1^{\rm exp.}=2.00\pm0.59,\;\;\; R_2^{\rm exp.}=1.15\pm0.34.
\end{eqnarray}
The difference between the two channels in the ratio $R_1$ is the
tree operator and electroweak penguin operators. Since the
contribution of tree operator is smaller than QCD penguins and  the
two kinds of amplitudes are perpendicular with each other due to the
CKM angle $\gamma$  close to $90^\circ$, the tree operator can not
change the branching ratio of $\bar B^0\to a_1^+K^-$ too much. Thus
this does not improve theoretical predictions on
$R_1$. 
Large electro-weak penguins may help us to diminish the large
deviation for $R_1$. In the $\bar B^0\to b_1^+K^-$ and $B^-\to b_1^0
K^-$ decays, the factorization formulae are exactly the same since
the $b_1$ decay constant vanishes. Thus in order to explain the
large ratio $R_2$, one needs some mechanism beyond factorization to
enhance the ratio of $R_2$ by roughly 2.5.

In the above, we have analyzed the charmless non-leptonic $B\to AP$
data under the factorization approach. The decay constant of $a_1$
meson and $B\to a_1,b_1$ form factors $V_0$ are extracted from the
$\bar B^0\to a_1\pi$ ad $\bar B^0\to b_1\pi$ decays. The form
factors are consistent with the predictions evaluated in
light-cone-sum-rules and the PQCD approach. But there exist several
problems which can be summarized as:
\begin{itemize}

\item The Wilson coefficient combination $a_2$ needs to be enhanced to
$a_2=0.5$ in order to solve the problem in $B^-\to a_1^-\pi^0$ and
$B^-\to a_1^0\pi^-$.

\item Since the form factor $B\to a_1$ and $B\to b_1$ are almost
equal in magnitude, the $\bar B^0\to a_1^+K^-$ and $\bar B^0\to
b_1^+K^-$ decays should possess similar and large branching ratios.
Compared with the experimental data, theoretical predictions needs
to be reduced by the factors of 2.8 and 5.5, respectively.

\item $B^-\to a_1^-\bar K^0$ and $B^-\to b_1^0K^-$ are related to $\bar B^0\to (a_1^+,b_1^+)K^-$
through relations given in Eq.~\eqref{eq:ratios} which also have
large deviations from the data.
\end{itemize}

\section{The soft-collinear effective theory}

The recent development  of SCET makes the analysis of $B\to M_1M_2$
decays on a more rigorous foundation.  The SCET is a powerful method
to systematically separate the dynamics at different scales: hard
scale $m_b$ ($b$ quark mass), hard intermediate scale
$\mu_{hc}=\sqrt{m_b\Lambda_{QCD}}$, soft scale and to sum large logs
using the renormalization group technics. Integrating out the hard
fluctuations, we arrive at the intermediate effective theory-
SCET$_{I}$   where the factorization formulae for $B\to M_1M_2$
decays to leading power in $\lambda\equiv\sqrt {\Lambda_{QCD}/m_b}$
are given by:
\begin{eqnarray}
 {\cal A}(B\to M_1M_2)&=&\frac{G_F}{\sqrt 2}m_B^2\left\{ f_{M_1} \int du \phi_{M_1}(u) T_1(u) \zeta^{B\to M_2}
 \right.\nonumber\\
   && \;\;\;\;\left.   +f_{M_1} \int du \phi_{M_1}(u) \int dz T_{1J}(u,z) \zeta_J^{B\to M_2}(z)+(1\leftrightarrow
   2)\right\},
\end{eqnarray}
where functions $\zeta$ and $\zeta_J$ also enter into the
heavy-to-light form factors. $T_1(u)$ and $T_{1J}(u,z)$ are hard
kernels which can be calculated using perturbation theory. With the
hard-collinear fluctuation integrated out, the final effective
theory-SCET$_{II}$ is obtained where the function $\zeta_J$ can be
factorized into convolutions of LCDAs with hard kernels:
\begin{eqnarray}
 \zeta_J^{B\to M_2}(z)=\phi_B(\omega)\otimes
 J(z,\omega,v)\otimes\phi_{M_2}(v).
\end{eqnarray}
$J(z,\omega,v)$ is the hard kernel and $\phi_B$ and $\phi_{M_1,M_2}$
are the light-cone distribution amplitudes (LCDAs). With our
knowledge on these LCDAs, one can predict the decay amplitude by
convoluting the LCDAs with the perturbatively calculated hard
kernels. But there is another alternative way for phenomenological
studies: one can fit experimental results, including branching
ratios and CP asymmetries, to determine essential non-perturbative
inputs. Note that in this way, no expansions in $\alpha_s(\sqrt
{m_b\Lambda_{QCD}})$ are needed and thus the exploration of the
convergence is spontaneously avoided. This method is especially
useful at tree level: $T_1(u)$ is a constant and $T_{1J}(u,z)$ is a
function of one argument $u$. It leads to a rather simple form for
decay amplitudes:
\begin{eqnarray}
 {\cal A}(B\to M_1M_2)&=&\frac{G_F}{\sqrt 2}m_B^2\left\{ f_{M_1}  T_1 \zeta^{B\to M_2}
 +f_{M_1} \int du \phi_{M_1}(u)  T_{1J}(u) \zeta_J^{B\to M_2} +(1\leftrightarrow 2)\right\},
\end{eqnarray}
where the functions $\zeta^{B\to M_2}$ and
\begin{eqnarray}
 \zeta_J^{B\to M_2}=\int dz \zeta_J^{B\to M_2}(z)
\end{eqnarray}
are treated as non-perturbative parameters to be fitted from the
data. With the help of the flavor SU(3) symmetry, the $B\to AP$
decays  involve only 6 parameters:
\begin{eqnarray}
 \zeta^{B\to P},\;\;\; \zeta_J^{B\to P},\;\;\;\zeta^{B\to ^1P_1},\;\;\; \zeta_J^{B\to ^1P_1},\;\;\;
 \zeta^{B\to ^3P_1},\;\;\; \zeta_J^{B\to ^3P_1},
\end{eqnarray}
 which contribute to the $B\to P$
and $B\to A$ form factors.

Including the non-perturbative contributions from loop diagrams
involving $c\bar
c$~\cite{Bauer:2004tj,Bauer:2005wb,Bauer:2005kd,Colangelo:1989gi,Ciuchini:1997hb,Ciuchini:2001gv},
the SCET  can successfully explain most of  $B\to PP$ and $B\to VP$
decays~\cite{Bauer:2005kd,Williamson:2006hb,Wang:2008rk}. This
phenomenological approach has many important features. In $b\to d$
transitions such as $\bar B^0\to \pi^+\pi^-$, tree operators provide
the dominant contributions and contributions from charming penguins
and penguin operators are sub-leading. From the experience in $B\to
PP$ and $B\to VP$ phenomenological study, we know that the
hard-scattering form factor $\zeta_J$ is potentially large.
Furthermore, as we have shown in Ref.~\cite{Wang:2008rk}, the
corresponding Wilson coefficient is of order 1 which amounts to a
large effective Wilson coefficient $a_2$. Here we take $\bar B^0\to
a_1^-\pi^+$ and $B^-\to a_1^-\pi^0$ as an example: if
hard-scattering form factors are equal with soft form factors for
pion and $a_1$ meson: $\zeta=\zeta_J$, the effective Wilson
coefficient equals to $a_2\simeq \frac{\zeta_J}{\zeta+\zeta_J}=0.5$.
Thus it is easy to solve the problems in $B\to a_1\pi$ decays under
the SCET framework.

For decays induced by $b\to s$ transition, since tree operators are
suppressed by the CKM matrix elements
$|V_{ub}V_{us}^*/(V_{cb}V_{cs}^*)|\sim 0.02$ and penguin operators
have smaller   Wilson coefficients (${\rm max}[C_{3-10}]\ll
\alpha_s(2m_c)C_1$), charming penguins play a significant role. Due
to the non-perturbative nature, charming penguins are totally
unknown from perturbation theory and needs to be extracted from
data. This stuff depends on the three involved mesons: $B$ meson,
recoiling meson and emitted meson. Thus in order to predict physical
observables, too many parameters for charming penguins are required.
An efficient way to reduce the independent inputs is to utilize the
flavor SU(3) symmetry and as a result only 8 parameters for charming
penguins in $B\to AP$ decays are left. But even so, due to the lack
of data, one can always obtain proper branching ratios of $\bar
B^0\to b_1^+K^-$ and $\bar B^0\to a_1^+K^-$ by adjusting charming
penguins. Despite of that, there is another deficit: since the
inputs, form factors and charming penguins, have been assumed to
respect the SU(3) symmetry, large deviations of the ratios shown in
Eqs.~\eqref{eq:ratiosthe} and \eqref{eq:ratiosexp} can not be
eliminated by the SCET either.

\section{The perturbative QCD approach}

There is another commonly-accepted approach to handle hadronic $B$
decays: the perturbative QCD
approach~\cite{Keum:2000ph,Keum:2000wi,Lu:2000em}. The basic idea of
the PQCD approach is that it takes into account the transverse
momentum of the valence quarks in hadrons. Decay amplitudes and form
factors can be written as convolutions of wave functions with
perturbatively hard kernels integrated over the longitudinal and
transverse component. When considering radiative corrections, one
encounters double logarithm divergences when soft and collinear
momenta overlap. These large double logarithm can be resummed into
the Sudakov factor. Loop corrections to the weak decay vertex also
give rise to double logarithms in the threshold region. Resummation
of this type of double logarithms leads to the Sudakov factor $S_t$.
This factor decreases faster than any power of ${x}$ as
${x\rightarrow 0}$ and changes the behavior at the end-point region.
The Sudakov factor and threshold resummation make the PQCD approach
more self-consistent. This approach have successfully explained the
$B\to \pi\pi$ and $B\to \pi K$ decay rates  and CP
asymmetries~\cite{Hong:2005wj} together with the proper
polarizations in $B\to VV$ decays \cite{vv}.

In the PQCD approach, the predicted $B\to a_1$ form
factor~\cite{Wang:2007an} is consistent with the one derived from
the data, thus our PQCD prediction on ${\cal BR}(\bar B^0\to
a_1^+\pi^-)$ is in good agreement with the data. But due to the
small value of $a_2$, the color-suppressed contribution is too small
to explain the large decay rates of $B^-\to a_1^-\pi^0$ and $B^-\to
a_1^0\pi^-$. The investigations of next-to-leading order corrections
in Ref.~\cite{Li:2006cva} show that the branching ratio of $B^-\to
\pi^-\pi^0$ is enhanced by the factor $4.0/3.5$ while $\bar B^0\to
\pi^+\pi^-$ is reduced by $6.5/7.0$.  But even if we assume the same
$k$ factor for $B\to a_1\pi$ decays, the PQCD predictions on $B^-\to
a_1^-\pi^0$ and $B^-\to a_1^0\pi^-$ are still smaller than the data.
The PQCD prediction on the $B\to b_1$ form factor is large, thus the
branching ratio of $\bar B^0\to b_1^+\pi^-$ is 2 times larger than
the experimental data and the QCDF results.
From the factorization formulae of $B\to PP$ decays given in the
literature \cite{Lu:2000em}, one can see that the  contributions
from hard spectator scattering diagrams are small due to the
cancelation between two diagrams where a gluon is attached to either
the positive quark or the anti-quark in the emitted hadron. But if
the emitted meson is a $P$-wave meson and the twist-2 LCDA is
anti-symmetric (like a scalar or an axial-vector meson with quantum
number $^{2S+1}L_J=^1P_1$), the two diagrams give constructive
contributions to make them sizable. For example, the large hard
spectator scattering contributions to $B^-\to b_1^0\pi^-$  make
${\cal BR}(B^-\to b_1^0\pi^-)>\frac{1}{2}{\cal BR}(\bar B^0\to
b_1^+\pi^-)$. Moreover, annihilation diagrams play an important role
in the PQCD approach which often enters into decay amplitudes as
imaginary. It provides the dominant strong phase which are essential
to explain the large CP asymmetries.  Thus unlike the situation in
the QCDF approach, annihilation diagrams do not cancel with emission
diagrams in $\bar B^0\to b_1^+K^-$ which results in much larger
predictions on branching ratios of $\bar B^0\to b_1^+K^-$. Similar
as the factorization approach,  there are large differences between
the PQCD approach predictions on ratios $R_{1,2}$ and those
extracted from the data.

\begin{table}
\caption{Theoretical predictions and experimental results
\cite{Aubert:2006dd,:2007jn,Aubert:2006gb,Palombo:2007jj,Aubert:2007xd,:2007kp,Aubert:2007ds,Mohanty:2007pj,:2008eq}
on branching ratios (in unit of $10^{-6}$) of $B\to a_1(b_1)\pi(K)$
decays. The QCDF predictions are quoted from
Ref.~\cite{Cheng:2007mx}. In the PQCD approach, the uncertainties
are from: (i) the hadronic inputs: decay constants of $B$ meson, and
shape parameters in the wave function of $B$ meson; (ii)
$\Lambda_{QCD}$, the hard scale $t$ and the threshold resummation
parameter $c$; (iii) the CKM matrix elements $V_{ub}$ and $\gamma$
angle. In the SCET framework,  the uncertainties are from: (i)
hadronic parameters: form factors and charming penguins; (ii) the
CKM matrix elements. } \label{tab:BR}
\begin{tabular}{|c|c|c|c|c|c|}
  \hline\hline
 channel & QCDF & PQCD & SCET & Exp.  \\\hline
 $B^-\to a_1^- \pi^0$   & $14.4^{+1.4+3.5+2.1}_{-1.3-3.2-1.9}$ & $8.1^{+4.1+2.1+0.7}_{-2.7-1.2-0.9}$
                        & $19.0_{-4.7-1.7}^{+5.1+1.8}$ &$26.4\pm5.4\pm4.1$\\
 $B^-\to a_1^0 \pi^-$   & $7.6^{+0.3+1.7+1.4}_{-0.3-1.3-1.0}$  & $6.7_{-2.2-1.7-0.7}^{+2.9+2.8+0.5}$
                        & $17.2^{+4.7+1.7}_{-4.3-1.6}$ & $20.4\pm4.7\pm3.4$    \\
 $\overline B^0\to a_1^-\pi^+$       & $23.4^{+2.3+6.2+1.9}_{-2.2-5.5-1.3}$& $15.7_{-5.6-3.6-1.7}^{+8.3+5.9+1.2}$
                        & $17.0^{+5.8+1.6}_{-5.2-1.4}$& $21.0\pm5.4$  \\
 $\overline B^0\to a_1^+ \pi^-$  & $9.1^{+0.2+2.2+1.7}_{-0.2-1.8-1.1}$ & $12.7_{-4.4-3.8-1.3}^{+5.6+6.2+0.9}$
                        &  $10.7^{+2.5+1.0}_{-2.4-0.9}$ &$12.2\pm4.5$  \\
 $B^0/\overline B^0\to a_1^+ \pi^-$  & ---  &  $28.1_{-9.9-7.3-3.0}^{+13.8+12.0+2.1}$
                        &  $28.2_{-5.9-2.4}^{+6.5+2.6}$ & \\
 $B^0/\overline B^0\to a_1^- \pi^+$  & --- &  $28.6_{-10.1-7.4-3.0}^{+13.9+12.1+2.2}$
                        &   $27.1_{-6.2-2.3}^{+6.9+2.5}$& \\
 $\overline B^0\to a_1^\pm \pi^\mp$   & $32.5^{+2.5+8.4+3.6}_{-2.4-7.3-2.4}$ & $28.3_{-10.0-7.4-3.0}^{+13.9+12.0+2.2}$
                        & $27.7_{-5.7-2.3}^{+6.3+2.5}$&$31.7\pm3.7$\\
 $\overline B^0\to a_1^0 \pi^0$   & $0.9^{+0.1+0.3+0.7}_{-0.1-0.2-0.3}$ & $0.12_{-0.04-0.03-0.02}^{+0.07+0.02+0.02}$
                        & $5.5_{-1.5-0.6}^{+1.7+0.6}$&   \\\hline
 $B^-\to a_1^0  K^-$  & $13.9^{+0.9+9.5+12.9}_{-0.9-5.1-~4.9}$    & $15.4_{-5.4-~5.5-2.5}^{+7.8+10.1+2.4}$
                        & $10.5_{-2.9-1.5}^{+3.3+1.8}$& \\
 $B^-\to a_1^- \overline K^0$   & $21.6^{+1.2+16.5+23.6}_{-1.1-~8.5-11.9}$  & $25.5_{-~9.2-10.2-3.9}^{+12.9+18.0+3.7}$
                        & $15.5_{-5.0-2.1}^{+5.8+2.5}$ & $34.9\pm 5.0\pm4.4$   \\
 $\overline B^0\to a_1^+ K^-$   & $18.3^{+1.0+14.2+21.1}_{-1.0-~7.2-~7.5}$  &$20.6_{-7.3-8.5-3.3}^{+10.2+14.6+3.2}$
                        &  $15.8_{-4.9-2.3}^{+5.6+2.7}$&  $16.3\pm2.9\pm2.3$  \\
 $\overline B^0\to a_1^0 \overline K^0$   & $6.9^{+0.3+6.1+9.5}_{-0.3-2.9-3.2}$ &$8.0_{-2.8-3.4-1.2}^{+3.9+6.4+1.2}$
                        & $6.3_{-2.1-0.8}^{+2.5+1.0}$ & \\\hline\hline
 $B^-\to b_1^- \pi^0$   & $0.4^{+0.0+0.2+0.4}_{-0.0-0.1-0.2}$    & $1.0^{+0.2+0.3+0.1}_{-0.2-0.2-0.2}$
                        & $2.0^{+0.8+0.2}_{-0.6-0.2}$ & $<3.3$ \footnote{The experimental data~\cite{:2008eq} is obtained
                        on the assumption
                        that the daughter decay $b_1\to\pi\omega$ has a branching ratio ${\cal BR}=1$.}\\
 $B^-\to b_1^0 \pi^-$   & $9.6^{+0.3+1.6+2.5}_{-0.3-1.6-1.5}$    & $5.1_{-1.9-1.7-0.5}^{+3.1+3.1+0.3}$
                        &  $5.0_{-1.2-0.4}^{+1.3+0.5}$& $6.7\pm1.7\pm1.0$ \\
 $\overline B^0\to b_1^-\pi^+$    & $0.3^{+0.1+0.1+0.3}_{-0.0-0.1-0.1}$     & $1.4_{-0.4-0.2-0.2}^{+0.4+0.1+0.1}$
                        & $0.6_{-0.2-0.1}^{+0.3+0.1}$&\\
 $\overline B^0\to b_1^+ \pi^-$   & $11.2^{+0.3+2.8+2.2}_{-0.3-2.4-1.9}$    & $18.7_{-6.4-4.5-1.9}^{+9.6+8.2+1.3}$
                        &  $7.7_{-1.9-0.7}^{+2.1+0.7}$ & \\
 $B^0/\overline B^0\to b_1^+ \pi^-$  & ---  & $14.8_{-5.6-3.8-1.7}^{+8.5+6.6+1.3}$
                        & $5.0_{-1.5-0.5}^{+1.8+0.6}$& \\
 $B^0/\overline B^0\to b_1^- \pi^+$  &---   &  $25.6_{-8.3-6.0-2.6}^{+11.4+9.6+1.6}$
                        & $11.6_{-2.5-0.9}^{+2.7+1.0}$ & \\
 $\overline B^0\to b_1^\pm \pi^\mp$& $11.4^{+0.4+2.9+2.5}_{-0.3-2.5-2.0}$   & $20.2_{-6.9-4.9-2.1}^{+9.9+8.1+1.4}$
                        & $8.3_{-1.9-0.7}^{+2.1+0.7}$& $10.9\pm1.2\pm0.9$    \\
 $\overline B^0\to b_1^0 \pi^0$    & $1.1^{+0.2+0.1+0.2}_{-0.2-0.1-0.2}$    & $1.5_{-0.5-0.3-0.2}^{+0.6+0.3+0.1}$
                        & $1.8_{-0.4-0.1}^{+0.5+0.2}$ & $<1.9 ~^a$\\\hline
 $B^-\to b_1^0  K^-$    & $6.2^{+0.5+5.0+6.4}_{-0.5-2.5-5.2}$    & $24.9_{-7.8-9.3-3.9}^{+9.8+14.9+3.7}$
                        & $4.6_{-1.5-0.6}^{+1.9+0.7}$& $9.1\pm1.7\pm1.0$ \\
 $B^-\to b_1^- \overline K^0$      & $14.0^{+1.3+11.5+13.9}_{-1.2-~5.9-~8.3}$ &$55.0_{-17.0-21.2-8.3}^{+23.6+33.5+8.0}$
                        & $8.6_{-3.1-1.2}^{+3.8+1.4}$& $9.6\pm1.7\pm0.9 ~^a$ \\
 $\overline B^0\to b_1^+ K^-$      & $12.1^{+1.0+9.7+12.3}_{-0.9-4.9-30.2}$ & $42.9_{-13.4-16.9-6.9}^{+17.7+26.9+6.6}$
                        & $8.5_{-2.8-1.1}^{+3.5+1.3}$ & $7.4\pm1.0\pm1.0$  \\
 $\overline B^0\to b_1^0 \overline K^0$& $7.3^{+0.5+5.4+6.7}_{-0.5-2.8-6.5}$& $23.3_{-6.8-8.8-3.6}^{+10.6+15.5+3.5}$
                        & $4.0_{-1.4-0.6}^{+1.8+0.7}$&  $<7.8 ~^a$\\\hline\hline
\end{tabular}
\end{table}

\section{Numerical Results}

 In the PQCD framework
and SCET framework, we   calculate the decay rates, direct CP
asymmetries and time-dependent CP asymmetries shown in table
\ref{tab:BR},\ref{tab:Acp},\ref{tab:timecpapibpi} and
\ref{tab:timecpneutral}. We have adopted the same conventions with
Ref.~\cite{Cheng:2007mx} for observables in time-dependent decay
widths of $B\to a_1^\pm\pi^\mp$ and $B\to
b_1^\pm\pi^\mp$~\footnote{The ${\cal A}_{a_1^\pm\pi^\mp}$ in the
present paper correspond to ${\cal A}_{a_1^\mp\pi^\pm}$ defined in
Ref.~\cite{Barberio:2007cr,Charles:2004jd,Bona:2005vz}.}.
 In the SCET calculation, we use the following
values for the 14 inputs:
\begin{eqnarray}\label{eq:SCETinputs}
 \zeta^{B\to\pi}&=&0.12,\;\;\;\zeta^{B\to\pi}_J=0.12,\nonumber\\
 \zeta^{B\to a_1}&=&0.17,\;\;\;\zeta^{B\to a_1}_J=0.17,\nonumber\\
 \zeta^{B\to b_1}&=&-0.16,\;\;\;\zeta^{B\to b_1}_J=-0.16,\nonumber\\
 |A_{cc}^{^3P_1P}|&=&40\times 10^{-4},\;\;\;
 arg[A_{cc}^{^3P_1P}]=160^\circ,\nonumber\\
 |A_{cc}^{P^3P_1}|&=&40\times 10^{-4},\;\;\;
 arg[A_{cc}^{P^3P_1}]=145^\circ,\nonumber\\
 |A_{cc}^{^1P_1P}|&=&40\times 10^{-4},\;\;\;
 arg[A_{cc}^{^1P_1P}]=155^\circ,\nonumber\\
 |A_{cc}^{P^1P_1}|&=&35\times 10^{-4},\;\;\;
 arg[A_{cc}^{P^1P_1}]=100^\circ.
\end{eqnarray}
We should point out that this set of inputs is presented by hand
instead of any reasonable way. To test the sensitivities on these
parameters, we show the first uncertainty in numerical results by
varying the form factors by $0.03$, $20\%$ for magnitudes of
charming penguins and $20^\circ$ for the phases. The second
uncertainty is from CKM matrix elements. In the PQCD calculation, we
have used the same inputs as those in
Ref.~\cite{Wang:2007an,Ali:2007ff,Li:2008ts}. The theoretical
uncertainties are from: (i) the hadronic inputs: decay constants of
$B$ meson, and shape parameters in the wave function of $B$ meson;
(ii) $\Lambda_{QCD}$, the hard scale $t$ and the threshold
resummation parameter $c$; (iii) the CKM matrix elements $V_{ub}$
and $\gamma$ angle. The factorization formula for each type of
diagrams in $B\to AP$ decays are the same with those in $B\to PP$
decays which can be found in the literature~\footnote{There still
exist two differences between the factorizable emission diagrams of
$B\to AP$ and $B\to PP$ decays: the axial-vector meson can not be
generated by the scalar or pseudo-scalar current, thus the chiraly
enhanced penguins vanish; due to the vanishing decay constant, $b_1$
can not be factorized from the $B$ meson and the recoiling meson.}.
Because of the same flavor structures, the hard spectator scattering
diagrams often accompany with the factorizable diagrams. One only
needs to consider the flavor structure for factorizable diagrams and
to use meson matrices   by evaluating the master
equations~\cite{Beneke:2003zv}. For the CKM matrix elements, we use
the updated global fit results from CKMfitter group
\cite{Charles:2004jd}:
\begin{eqnarray}
 &&V_{ud}=0.97400,\;\;\;\;\; V_{us}=0.22653,\;\;\; |V_{ub}|=(3.57^{+0.17}_{-0.17})\times
 10^{-3},\nonumber\\
 && V_{cd}=-0.22638 ,\;\;\; V_{cs}=0.97316,\;\;\; V_{cb}=(40.5^{+3.2}_{-2.9})\times
 10^{-3},\nonumber\\
 &&\beta=(21.7^{+0.017}_{-0.017})^\circ,\;\;\; \gamma=(67.6^{+2.8}_{-4.5})^\circ.
\end{eqnarray}

\begin{table}
\caption{Similar as table I. but for direct CP asymmetries (in $\%$)
of $B\to a_1(b_1)\pi(K)$ decays. }\label{tab:Acp}
\begin{tabular}{|c|c|c|c|c|c|}
  \hline\hline
 channel & QCDF & PQCD & SCET & Exp.  \\\hline
 $B^-\to a_1^- \pi^0$   & $0.5^{+0.3+0.6+12.0}_{-0.2-0.3-11.0}$                 & $1.6_{-0.6-1.3-0.1}^{+0.0+0.1+0.2}$
                        & $-5.4_{-10.1-0.5}^{+10.7+0.5}$&  \\
 $B^-\to a_1^0 \pi^-$   & $-4.3^{+0.3+1.4+14.1}_{-0.3-2.2-14.5}$                &$-0.9_{-0.3-0.3-0.1}^{+0.6+0.3+0.1}$
                        & $5.7_{-11.3-0.5}^{+11.1+0.5}$ & \\
 $\overline B^0\to a_1^+ \pi^-$ &$-3.6^{+0.1+0.3+20.8}_{-0.1-0.5-20.2}$ & $12.6_{-1.2-2.5-1.1}^{+1.8+3.5+1.0}$
                        &  $21.5_{-12.7-1.9}^{+11.3+1.6}$  & $7\pm21\pm15$   \\
 $\overline B^0\to a_1^-\pi^+$   & $-1.9\pm0.0\pm0.0^{+14.6}_{-14.3}$      &  $11.7_{-1.9-2.0-1.1}^{+2.1+2.7+1.1}$
                        & $10.4_{-10.6-0.9}^{+~9.9+0.8}$ &  $15\pm15\pm7$   \\
 $\overline B^0\to a_1^0 \pi^0$   & $60.1^{+4.6+6.8+37.6}_{-4.9-8.3-60.7}$ & $28.9_{-22.1-88.1-2.5}^{+~7.6+42.5+2.6}$
                        & $-29.5_{-13.0-2.8}^{+15.7+2.6}$&   \\\hline
 $B^-\to a_1^- \overline K^0$   & $0.8^{+0.0+0.1+0.6}_{-0.0-0.1-0.0}$            &$-1.0_{-0.0-0.2-0.1}^{+0.2+0.2+0.1}$
                        & $0.3_{-0.2-0.0}^{+0.2+0.0}$ & $12\pm 11\pm2$   \\
 $B^-\to a_1^0  K^-$  & $8.4^{+0.3+1.4+10.3}_{-0.3-1.6-12.0}$                    &$-6.1_{-1.3-1.4-0.6}^{+1.1+1.3+0.6}$
                        & $-25.6_{-14.8-2.4}^{+14.9+2.3}$ & \\
 $\overline B^0\to a_1^+ K^-$   & $2.6^{+0.0+0.7+10.1}_{-0.1-0.7-11.0}$          &$-8.9_{-2.3-2.4-0.9}^{+1.5+2.1+0.8}$
                        & $-17.7_{-10.0-1.7}^{+10.3+1.6}$ &  $-16\pm12\pm1$  \\
 $\overline B^0\to a_1^0 \overline K^0$   & $-7.7^{+0.6+2.1+6.8}_{-0.6-2.2-7.0}$ &$-1.8_{-0.3-0.6-0.2}^{+0.3+0.6+0.2}$
                        & $17.9_{-11.1-1.6}^{+10.1+1.4}$& \\
 \hline\hline
 $B^-\to b_1^- \pi^0$   & $-36.5^{+4.4+18.4+82.2}_{-4.3-17.7-59.6}$   & $12.5_{-27.3-40.3-0.6}^{+16.0+23.5+1.1}$
                        & $34.1_{-26.7-2.9}^{+27.1+3.2}$ & $5\pm16\pm2$ \\
 $B^-\to b_1^0 \pi^-$   & $0.9^{+0.6+2.3+18.0}_{-0.4-2.7-20.5}$       & $-65.8_{-9.0-8.2-3.8}^{+11.1+13.4+4.6}$
                        &$-17.7_{-14.6-1.3}^{+13.8+1.6}$ &    \\
 $\overline B^0\to b_1^+ \pi^-$    & $-4.0^{+0.2+0.4+26.2}_{-0.0-0.6-25.5}$& $-25.0_{-4.3-4.1-1.9}^{+4.0+3.9+2.2}$
                        & $-42.3_{-9.4-3.5}^{+9.3+3.6}$ &\\
 $\overline B^0\to b_1^-\pi^+$     &$66.1^{+1.2+7.4+30.3}_{-1.4-4.8-96.6}$ & $49.0_{-8.0-6.0-4.0}^{+3.5+7.1+4.2}$
                        &  $0$ & \\
 $\overline B^0\to b_1^0 \pi^0$    & $53.4^{+6.4+9.0+5.2}_{-6.3-7.3-4.7}$        & $15.9_{-7.8-10.7-1.4}^{+4.0+7.2+1.0}$
                        & $52.7_{-13.9-3.9}^{+12.7+2.7}$ &\\\hline
 $B^-\to b_1^- \overline K^0$      & $1.4^{+0.1+0.1+5.6}_{-0.1-0.1-0.1}$ & $-0.30_{-0.38-0.58-0.03}^{+0.02+0.00+0.03}$
                        & $-0.8_{-0.2-0.1}^{+0.2+0.1}$ & $-3\pm15\pm2$\\
 $B^-\to b_1^0  K^-$    & $18.7^{+1.6+7.8+57.7}_{-1.7-6.1-44.9}$      & $19.4_{-0.4-4.0-1.8}^{+0.0+4.4+1.8}$
                        & $46.3_{-8.8-3.9}^{+10.5+3.8}$ & $-46\pm20\pm2$\\
 $\overline B^0\to b_1^+ K^-$ & $5.5^{+0.2+1.2+47.2}_{-0.3-1.2-30.2}$       & $16.6_{-2.3-3.2-1.5}^{+2.4+3.8+1.6}$
                        & $46.3_{-8.8-3.9}^{+10.5+3.8}$ & $-7\pm12\pm2$  \\
 $\overline B^0\to b_1^0 \overline K^0$& $-8.6^{+0.8+3.3+8.3}_{-0.8-4.2-25.4}$   & $-4.3_{-1.6-1.8-0.4}^{+1.6+1.8+0.4}$
                        & $-0.8_{-0.2-0.1}^{+0.2+0.1}$ &  \\
  \hline\hline
\end{tabular}
\end{table}

Predictions in the QCDF approach are also collected in the tables to
make a comparison~\cite{Cheng:2007mx} In the QCDF approach, $a_2$
(to be precise, $\alpha_2$) is much smaller than 0.5, thus their
amplitude from color-suppressed tree diagrams is not large enough to
resolve the problem in $B^0/\bar B^0\to a_1^\pm\pi^\mp$ and $B^0\to
(a_1^-\pi^0,a_1^0\pi^-)$ decays. Their prediction on the branching
ratio of $\bar B^0\to a_1^+K^-$ is compatible with the data. For
$B\to b_1K$, they found that decay rates are sensitive to the
interference between emission diagrams and annihilation diagrams.
The small decay rate of $\bar B^0\to b_1^+K^-$ arises from the
destructive interference between emission diagrams and
annihilations, thus the prediction on branching ratio $\bar B^0\to
b_1^+K^-$ is basically consistent with the data. But their
predictions on four ratios of branching fractions $R_{1-4}$ ($R_3$
and $R_4$ are related to ratios $R_1$ and $R_2$ defined in the
present paper; their ratios $R_1$ and $R_2$ characterize the
magnitude of color-suppressed contributions in $B\to a_1\pi$ decay
modes.) deviate from experimental data.

Several remarks on the numerical results in the PQCD approach and
SCET approach are in order:
\begin{itemize}

\item The predictions on ${\cal BR}(\bar B^0\to a_1^-\pi^+)$ in both approaches are a
bit smaller than experimental data, because the decay constant of
$f_{a_1}=0.238$ GeV~\cite{Yang:2007zt} is a bit smaller than that
extracted from the data.

\item As we expected, color-suppressed contributions to $B\to a_1\pi$ decays
are large in the SCET framework but small in the PQCD approach: SCET
predictions are much larger and consistent with the present data
within the uncertainties.

\item In the PQCD approach, $\bar B^0\to b_1^-\pi^+$ occur via the
so-called hard spectator scattering  diagrams,
despite of the zero decay constant of $b_1$. 
In $B^-\to b_1^0\pi^-$, the hard spectator scattering  diagrams
contributions (tree operators), with a $b_1^0$ meson emitted, are
sizable and cancel with color-allowed contribution where the pion is
emitted. Thus the branching ratio of $B^-\to b_1^0\pi^-$ is smaller
than one half of ${\cal BR}(\bar B^0\to b_1^+\pi^-)$.

\item In the SCET approach, $\bar B^0\to b_1^-\pi^+$ only receive
contributions from charming penguins  and correspondingly the direct
CP asymmetry in this channel is 0. The predicted branching ratio is
smaller than the PQCD prediction but larger than the QCDF
prediction.

\item In the SCET, the direct CP asymmetries in $\bar B^0\to a_1^+K^-$ and
$B^-\to a_1^0 K^-$ have the same sign and   similar size. Moreover,
their branching ratios obey the simple relation: ${\cal BR}(\bar
B^0\to a_1^+K^-)=2 {\cal BR}(B^-\to a_1^0 K^-)$. It is also similar
for $\bar B^0\to b_1^+K^-$ and $B^-\to b_1^0 K^-$: the direct CP
asymmetries are equal with each other; the branching ratios also
satisfy the relation ${\cal BR}(\bar B^0\to b_1^+K^-)=2 {\cal
BR}(B^-\to b_1^0 K^-)$, where the small deviation arises from the
different mass and decay width of $\bar B^0$ and $B^-$ meson.

\item As expected, the two ratios $R_1$ and $R_2$ are predicted with large deviations from the data:
\begin{eqnarray}
 &&R_1=1.16,\;\;\; R_2=0.54,\;\;\;\; {\rm PQCD}  \\
 && R_1=0.91,\;\;\; R_2=0.50.\;\;\;\; {\rm SCET}
\end{eqnarray}

\item Predictions on the observables in time-dependent decay width
of $B^0/\bar B^0\to a_1^\pm\pi^\mp$ and $B^0/\bar B^0\to
b_1^\pm\pi^\mp$ are basically consistent with the experimental data
except the $\Delta S$, the $\alpha_{eff}$ in $\bar B^0\to
a_1^\pm\pi^\mp$ and ${\cal A}_{b_1\pi}$. For $\bar B^0\to b_1^\pm
\pi^\mp$ decays, predictions on $\Delta C$ in the two approaches are
close to $-1$ and they are consistent with the QCDF
prediction~\cite{Cheng:2007mx} and the data.  In the SCET framework,
the angle $\alpha^+_{eff}(\bar B^0\to b_1^\pm \pi^\mp)$ is equal to
$\frac{\pi}{2}-\beta$ which is also a consequence of the vanishing
decay constant of $b_1$ meson.

\end{itemize}

\begin{table}
\caption{Same as table~\ref{tab:BR} but for Time-dependent CP
asymmetry parameters in $B^0/\bar B^0\to a_1^\pm\pi^\mp$ and
$B^0/\bar B^0\to b_1^\pm\pi^\mp$ decays.}\label{tab:timecpapibpi}
\begin{tabular}{|c|c|c|c|c|}\hline
 Observables & QCDF &PQCD &  SCET& Exp. \\ \hline
 ${\cal A}_{a_1\pi}$ & $\phantom{-}0.003^{\,+0.001\,+0.002\,+0.043}_{\,-0.002\,-0.003\,-0.045}$
                     &  $-0.009_{-0.002-0.003-0.001}^{+0.002+0.002+0.001}$
                     & $0.02_{-0.08-0.00}^{+0.08+0.00}$
                      & $-0.07\pm0.07\pm0.02$\\
 $C$ & $\phantom{-}0.02^{\,+0.00\,+0.00\,+0.14}_{\,-0.00\,-0.00\,-0.14}$
 &         $-0.12_{-0.02-0.03-0.01}^{+0.02+0.02+0.01}$
 & $-0.15_{-0.07-0.01}^{+0.08+0.01}$
 &$-0.10\pm0.15\pm0.09$ \\
$\Delta C$ &
$\phantom{-}0.44^{\,+0.03\,+0.03\,+0.03}_{\,-0.04\,-0.05\,-0.04}$
 &      $0.11_{-0.01-0.05-0.01}^{+0.03+0.06+0.01}$
 & $0.23_{-0.19-0.00}^{+0.18+0.00}$
 &$0.26\pm0.15\pm0.07$ \\
$S$ &
 $ -0.37^{\,+0.01\,+0.05\,+0.09}_{\,-0.01\,-0.08\,-0.16}$
 &     $-0.23_{-0.01-0.03-0.14}^{+0.02+0.03+0.09}$
 & $-0.45_{-0.06-0.11}^{+0.07+0.08}$
   &$0.37\pm0.21\pm0.07$ \\
 $\Delta S$ & $0.01^{\,+0.00\,+0.00\,+0.02}_{\,-0.00\, -0.00\,-0.02}$
 & $-0.03_{-0.01-0.01-0.00}^{+0.01+0.01+0.00}$
 &  $0.02_{-0.05-0.00}^{+0.04+0.00}$ &$-0.14\pm0.21\pm0.06$ \\
 $\alpha_{\rm eff}^+$  & $(97.2^{\, +0.3\, +1.0\, +4.7}_{\,-0.3\, -0.6\, -2.5})^\circ$
 &$(93.8_{-0.4-0.4-2.8}^{+0.4+0.7+4.4})^\circ$
 & $(103.5_{-2.5-2.6}^{+2.4+4.0})^\circ$ &  \\
 $\alpha_{\rm eff}^-$  & $(107.0^{\, +0.5\, +3.6\, +6.6}_{\,-0.5\, -2.3\, -3.7})^\circ$
 & $(99.8_{-0.7-1.4-2.7}^{+0.5+1.5+4.2})^\circ$
 & $(104.8_{-3.2-2.6}^{+2.7+4.0})^\circ$ &  \\
 $\alpha_{\rm eff}$  & $(102.0^{\, +0.4\, +2.3\, +5.7}_{\,-0.4\, -1.5\, -3.1})^\circ$
 & $(96.8_{-0.6-0.9-2.7}^{+0.4+1.0+4.3})^\circ$
 & $(104.2_{-2.0-2.6}^{+1.8+4.0})^\circ$ &  $(78.6\pm 7.3)^\circ$ \\
\hline
 ${\cal A}_{b_1\pi}$ &
 $-0.06^{\,+0.01\,+0.01\,+0.23}_{\,-0.01\,-0.01\,-0.23}$
 & $-0.27_{-0.05-0.04-0.02}^{+0.05+0.04+0.02}$ & $-0.39_{-0.08-0.03}^{+0.08+0.03}$
 & $-0.05\pm0.10\pm0.02$\\
 $C$ & $-0.03^{\,+0.01\,+0.01\,+0.06}_{\,-0.02\,-0.02\,-0.01}$
   &$-0.03_{-0.01-0.01-0.00}^{+0.01+0.01+0.00}$ & $0.07_{-0.03-0.01}^{+0.04+0.02}$
   &$0.22\pm0.23\pm0.05$   \\
$\Delta C$ & $-0.96^{\,+0.03\,+0.02\,+0.08}_{\,-0.03\,-0.03\,-0.01}$
 & $-0.87_{-0.02-0.04-0.01}^{+0.02+0.04+0.01}$ & $-0.83_{-0.07-0.03}^{+0.08+0.03}$
& $-1.04\pm0.23\pm0.08$ \\
$S$ &
 $ 0.05^{\,+0.03\,+0.02\,+0.15}_{\,-0.03\,-0.02\,-0.26}$ &
  $0.08_{-0.01-0.02-0.04}^{+0.00+0.02+0.06}$ & $-0.46_{-0.10-0.03}^{+0.14+0.03}$ &
  \\
 $\Delta S$ & $0.12^{\,+0.04\,+0.04\,+0.08}_{\,-0.03\, -0.04\,-0.09}$
  & $-0.24_{-0.02-0.07-0.01}^{+0.01+0.06+0.02}$& $-0.17_{-0.05-0.02}^{+0.06+0.03}$ &
   $$ \\
 $\alpha_{\rm eff}^+$
\footnote{One needs to be careful about the phase of the $B$-meson
decay amplitudes~\cite{Beneke:2003zv}. For example, the $\bar B^0\to
b_1^-\pi^+$ and $B^0\to b_1^+\pi^-$ decay amplitudes are determined
as:
\begin{eqnarray}
 A(\bar B^0\to b_1^-\pi^+)&=&
 V_{ub}V_{ud}^*T+V_{cb}V_{cd}^*P,\;\;\;\
 A(B^0\to b_1^+\pi^-)=-[V_{ub}^*V_{ud}T+V_{cb}^*V_{cd}P].
\end{eqnarray}.}
   & $(107.6^{\, +0.7\, +3.5\, +155.4}_{\,-0.2\, -4.9\, -~17.8})^\circ$
 &$(174.1_{-0.8-3.9-3.2}^{+0.0+2.5+5.0})^\circ$
 & $68.3^\circ$
 &\\
 $\alpha_{\rm eff}^-$  & $(101.3^{\, +0.4\, +2.1\, +4.9}_{\,-0.4\, -1.4\, -8.6})^\circ$
 & $(13.3_{-0.3-1.7-2.7}^{+0.2+1.9+4.1})^\circ$
 & $(3.4_{-0.1-2.9}^{+0.2+4.6})^\circ$
 & \\
 $\alpha_{\rm eff}$  & $(104.4^{\, +0.6\, +2.6\, +80.4}_{\,-0.3\, -2.1\, -~1.6})^\circ$
 & $(93.7_{-0.5-1.6-2.8}^{+0.0+0.9+4.5})^\circ$
 & $(35.8_{-0.0-1.5}^{+0.1+2.3})^\circ$
 & \\
 \hline
\end{tabular}
\end{table}

\begin{table}
\caption{Mixing-induced CP asymmetries in $\bar B^0\to a_1^0K_S$ and
$\bar B^0\to b_1^0K_S$ decays. 
  }\label{tab:timecpneutral}
\begin{tabular}{|c|c|c|c|}
  \hline\hline
 Channel    & PQCD & SCET      \\\hline
 $  \overline B^0\to a_1^0 \pi^0$    & $0.09_{-0.19-0.87-0.08}^{+0.20+0.78+0.08}$ & $0.48_{-0.14-0.15}^{+0.11+0.09}$ \\
 $\overline B^0\to a_1^0  K_S$       & $0.71^{+0.01+0.02+0.00}_{-0.01-0.02-0.00}$& $0.85_{-0.06-0.01}^{+0.05+0.01}$ \\
 $\overline B^0\to b_1^0 \pi^0$      & $0.67^{+0.02+0.09+0.09}_{-0.00-0.06-0.07}$ & $0.61_{-0.11-0.06}^{+0.09+0.09}$ \\
 $\overline B^0\to b_1^0 K_S$        & $-0.61_{-0.01-0.03-0.01}^{+0.01+0.03+0.01}$& $-0.69$
 \\
  \hline\hline
\end{tabular}
\end{table}

\section{Summary}

In summary, we have investigated the $B\to a_1(b_1)\pi(K)$  decays
under the factorization framework and find large differences between
theoretical predictions and experimental data. In tree dominated
processes $B\to a_1\pi$, large contributions from color-suppressed
tree diagrams are required. In $\bar B^0\to (a_1^+, b_1^+)K^-$
decays, theoretical results are larger than data by factors of 2.8
and 5.5 respectively, meanwhile ratios $R_1$ and $R_2$ defined in
Eq.~\eqref{eq:ratios} are too much larger too. In the PQCD
framework, the predicted decay rates of $B\to a_1^\pm\pi^\mp$ are
consistent with data. But the other problems can not be resolved.
The SCET approach has the potential to resolve the first two
problems: if large hard-scattering form factors are allowed,
theoretical predictions ${\cal BR}(B^-\to a_1^-\pi^0)$ and  ${\cal
BR}(B^-\to a_1^0\pi^-)$ are in good agreement with data; with the
help of charming penguins, large branching ratios of $\bar B^0\to
(a_1^+, b_1^+)K^-$ are also pulled down to the same magnitude with
the data. However, the two problems on ratios in $b\to s$
transitions remain in the present theoretical methods. These two
problems may indicate some new mechanism,  from the non-perturbative
contributions such as final state interactions or new physics
scenarios, which needs further study.

\section*{Acknowledgements}

This work is partly supported by National Natural Science Foundation
of China under the Grant Numbers 10735080, 10625525 and 10525523. We
would like to acknowledge H.Y. Cheng, Y.L. Shen and D.S. Yang for
valuable discussions.


\begin{thebibliography}{11}

\bibitem{Aubert:2006dd}
  B.~Aubert {\it et al.}  [BABAR Collaboration],
  Phys.\ Rev.\ Lett.\  {\bf 97}, 051802 (2006)
  [arXiv:hep-ex/0603050].

\bibitem{:2007jn}
  K.~Abe {\it et al.}  [Belle Collaboration],
  arXiv:0706.3279 [hep-ex].



\bibitem{Aubert:2006gb}
  B.~Aubert {\it et al.}  [BABAR Collaboration],
  Phys.\ Rev.\ Lett.\  {\bf 98}, 181803 (2007)
  [arXiv:hep-ex/0612050].


\bibitem{Palombo:2007jj}
  F.~Palombo,
  arXiv:hep-ex/0703005.

\bibitem{Aubert:2007xd}
  B.~Aubert {\it et al.}  [The BABAR Collaboration],
  Phys.\ Rev.\ Lett.\  {\bf 99}, 241803 (2007)
  [arXiv:0707.4561 [hep-ex]].

\bibitem{:2007kp}
  B.~Aubert {\it et al.}  [BABAR Collaboration],
  Phys.\ Rev.\ Lett.\  {\bf 99}, 261801 (2007)
  [arXiv:0708.0050 [hep-ex]].


\bibitem{Aubert:2007ds}
  B.~Aubert {\it et al.}  [BABAR Collaboration],
  arXiv:0709.4165 [hep-ex].

\bibitem{Mohanty:2007pj}
  G.~B.~Mohanty  [BABAR Collaboration],
  arXiv:0711.4956 [hep-ex].



\bibitem{:2008eq}
  B.~Aubert {\it et al.}  [BABAR Collaboration],
  arXiv:0805.1217 [hep-ex].



\bibitem{Chen:2005cx}
  C.~H.~Chen, C.~Q.~Geng, Y.~K.~Hsiao and Z.~T.~Wei,
  Phys.\ Rev.\  D {\bf 72}, 054011 (2005)
  [arXiv:hep-ph/0507012].


\bibitem{Laporta:2006uf}
  V.~Laporta, G.~Nardulli and T.~N.~Pham,
  Phys.\ Rev.\  D {\bf 74}, 054035 (2006)
  [Erratum-ibid.\  D {\bf 76}, 079903 (2007)]
  [arXiv:hep-ph/0602243].

\bibitem{Calderon:2007nw}
  G.~Calderon, J.~H.~Munoz and C.~E.~Vera,
  Phys.\ Rev.\  D {\bf 76}, 094019 (2007)
  [arXiv:0705.1181 [hep-ph]].
%
\bibitem{Yang:2007sb}
  K.~C.~Yang,
  Phys.\ Rev.\  D {\bf 76}, 094002 (2007)
  [arXiv:0705.4029 [hep-ph]].

\bibitem{Cheng:2007mx}
  H.~Y.~Cheng and K.~C.~Yang,
  Phys.\ Rev.\  D {\bf 76}, 114020 (2007)
  [arXiv:0709.0137 [hep-ph]].




\bibitem{Buchalla:1995vs}
  G.~Buchalla, A.~J.~Buras and M.~E.~Lautenbacher,
  Rev.\ Mod.\ Phys.\  {\bf 68}, 1125 (1996)
  [arXiv:hep-ph/9512380].

\bibitem{Barberio:2007cr}
  E.~Barberio {\it et al.}  [Heavy Flavor Averaging Group (HFAG)
                  Collaboration],
  arXiv:0704.3575 [hep-ex]. The updated results can be found at
  http://www.slac.stanford.edu/xorg/hfag.



\bibitem{Wang:2007an}
  W.~Wang, R.~H.~Li and C.~D.~Lu,
  arXiv:0711.0432 [hep-ph].


\bibitem{Yang:}
  K.~C.~Yang, unpublished;
 H.~Hatanaka and K.~C.~Yang,
  arXiv:0804.3198 [hep-ph].

\bibitem{Wang:2008bw}
  Z.~G.~Wang,
  arXiv:0804.0907 [hep-ph].
%
%
%



\bibitem{Bauer:2004tj}
  C.~W.~Bauer, D.~Pirjol, I.~Z.~Rothstein and I.~W.~Stewart,
  Phys.\ Rev.\  D {\bf 70}, 054015 (2004)
  [arXiv:hep-ph/0401188].



\bibitem{Bauer:2005kd}
  C.~W.~Bauer, I.~Z.~Rothstein and I.~W.~Stewart,
  Phys.\ Rev.\  D {\bf 74}, 034010 (2006)
  [arXiv:hep-ph/0510241].





\bibitem{Bauer:2005wb}
  C.~W.~Bauer, D.~Pirjol, I.~Z.~Rothstein and I.~W.~Stewart,
  Phys.\ Rev.\  D {\bf 72}, 098502 (2005)
  [arXiv:hep-ph/0502094].








\bibitem{Colangelo:1989gi}
  P.~Colangelo, G.~Nardulli, N.~Paver and Riazuddin,
  Z.\ Phys.\  C {\bf 45}, 575 (1990).



\bibitem{Ciuchini:1997hb}
  M.~Ciuchini, E.~Franco, G.~Martinelli and L.~Silvestrini,
  Nucl.\ Phys.\  B {\bf 501} (1997) 271
  [arXiv:hep-ph/9703353].

\bibitem{Ciuchini:2001gv}
  M.~Ciuchini, E.~Franco, G.~Martinelli, M.~Pierini and L.~Silvestrini,
  Phys.\ Lett.\  B {\bf 515}, 33 (2001)
  [arXiv:hep-ph/0104126].



\bibitem{Williamson:2006hb}
  A.~R.~Williamson and J.~Zupan,
  Phys.\ Rev.\  D {\bf 74}, 014003 (2006)
  [Erratum-ibid.\  D {\bf 74}, 03901 (2006)]
  [arXiv:hep-ph/0601214].


\bibitem{Wang:2008rk}
  W.~Wang, Y.~M.~Wang, D.~S.~Yang and C.~D.~Lu,
  arXiv:0801.3123 [hep-ph].



\bibitem{Keum:2000ph}
  Y.~Y.~Keum, H.~n.~Li and A.~I.~Sanda,
  Phys.\ Lett.\  B {\bf 504}, 6 (2001)
  [arXiv:hep-ph/0004004].

\bibitem{Keum:2000wi}
  Y.~Y.~Keum, H.~N.~Li and A.~I.~Sanda,
  Phys.\ Rev.\  D {\bf 63}, 054008 (2001)
  [arXiv:hep-ph/0004173].

\bibitem{Lu:2000em}
  C.~D.~Lu, K.~Ukai and M.~Z.~Yang,
  Phys.\ Rev.\  D {\bf 63}, 074009 (2001)
  [arXiv:hep-ph/0004213].


\bibitem{Hong:2005wj}
  B.~H.~Hong and C.~D.~Lu,
  Sci.\ China {\bf G49}, 357 (2006)
  [arXiv:hep-ph/0505020].

\bibitem{vv}J. Zhu, Y.L. Shen, C.D. Lu, Phys. Rev. D72, 054015 (2005);
H-n Li, Phys. Lett. B622, 68 (2005);  H.W. Huang, Phys. Rev. D73,
014011 (2006)

\bibitem{Li:2006cva}
  H.~n.~Li and S.~Mishima,
  Phys.\ Rev.\  D {\bf 73}, 114014 (2006)
  [arXiv:hep-ph/0602214].





\bibitem{Bona:2005vz}
  M.~Bona {\it et al.}  [UTfit Collaboration],
  JHEP {\bf 0507}, 028 (2005)
  [arXiv:hep-ph/0501199].



\bibitem{Charles:2004jd}
  J.~Charles {\it et al.}  [CKMfitter Group],
  Eur.\ Phys.\ J.\  C {\bf 41}, 1 (2005)
  [arXiv:hep-ph/0406184]. The updated results can be found at
  http://ckmfitter.in2p3.fr/.





\bibitem{Ali:2007ff}
  A.~Ali, G.~Kramer, Y.~Li, C.~D.~Lu, Y.~L.~Shen, W.~Wang and Y.~M.~Wang,
  Phys.\ Rev.\  D {\bf 76}, 074018 (2007)
  [arXiv:hep-ph/0703162].

\bibitem{Li:2008ts}
  R.~H.~Li, C.~D.~Lu and H.~Zou,
  arXiv:0803.1073 [hep-ph].


\bibitem{Beneke:2003zv}
  M.~Beneke and M.~Neubert,
  Nucl.\ Phys.\  B {\bf 675}, 333 (2003)
  [arXiv:hep-ph/0308039].


\bibitem{Yang:2007zt}
  K.~C.~Yang,
  Nucl.\ Phys.\  B {\bf 776}, 187 (2007)
  [arXiv:0705.0692 [hep-ph]].



\end{thebibliography}
\end{document}